


\font\bigbold=cmbx12
\font\ninerm=cmr9
\font\eightrm=cmr8
\font\sixrm=cmr6
\font\fiverm=cmr5
\font\ninebf=cmbx9
\font\eightbf=cmbx8
\font\sixbf=cmbx6
\font\fivebf=cmbx5
\font\ninei=cmmi9  \skewchar\ninei='177
\font\eighti=cmmi8  \skewchar\eighti='177
\font\sixi=cmmi6    \skewchar\sixi='177
\font\fivei=cmmi5
\font\ninesy=cmsy9 \skewchar\ninesy='60
\font\eightsy=cmsy8 \skewchar\eightsy='60
\font\sixsy=cmsy6   \skewchar\sixsy='60
\font\fivesy=cmsy5
\font\nineit=cmti9
\font\eightit=cmti8
\font\ninesl=cmsl9
\font\eightsl=cmsl8
\font\ninett=cmtt9
\font\eighttt=cmtt8
\font\tenfrak=eufm10
\font\ninefrak=eufm9
\font\eightfrak=eufm8
\font\sevenfrak=eufm7
\font\fivefrak=eufm5
\font\tenbb=msbm10
\font\ninebb=msbm9
\font\eightbb=msbm8
\font\sevenbb=msbm7
\font\fivebb=msbm5
\font\tensmc=cmcsc10


\newfam\bbfam
\textfont\bbfam=\tenbb
\scriptfont\bbfam=\sevenbb
\scriptscriptfont\bbfam=\fivebb
\def\Bbb{\fam\bbfam}

\newfam\frakfam
\textfont\frakfam=\tenfrak
\scriptfont\frakfam=\sevenfrak
\scriptscriptfont\frakfam=\fivefrak
\def\frak{\fam\frakfam}

\def\smc{\tensmc}


\def\eightpoint{%
\textfont0=\eightrm   \scriptfont0=\sixrm
\scriptscriptfont0=\fiverm  \def\rm{\fam0\eightrm}%
\textfont1=\eighti   \scriptfont1=\sixi
\scriptscriptfont1=\fivei  \def\oldstyle{\fam1\eighti}%
\textfont2=\eightsy   \scriptfont2=\sixsy
\scriptscriptfont2=\fivesy
\textfont\itfam=\eightit  \def\it{\fam\itfam\eightit}%
\textfont\slfam=\eightsl  \def\sl{\fam\slfam\eightsl}%
\textfont\ttfam=\eighttt  \def\tt{\fam\ttfam\eighttt}%
\textfont\frakfam=\eightfrak \def\frak{\fam\frakfam\eightfrak}%
\textfont\bbfam=\eightbb  \def\Bbb{\fam\bbfam\eightbb}%
\textfont\bffam=\eightbf   \scriptfont\bffam=\sixbf
\scriptscriptfont\bffam=\fivebf  \def\bf{\fam\bffam\eightbf}%
\abovedisplayskip=9pt plus 2pt minus 6pt
\belowdisplayskip=\abovedisplayskip
\abovedisplayshortskip=0pt plus 2pt
\belowdisplayshortskip=5pt plus2pt minus 3pt
\smallskipamount=2pt plus 1pt minus 1pt
\medskipamount=4pt plus 2pt minus 2pt
\bigskipamount=9pt plus4pt minus 4pt
\setbox\strutbox=\hbox{\vrule height 7pt depth 2pt width 0pt}%
\normalbaselineskip=9pt \normalbaselines
\rm}


\def\ninepoint{%
\textfont0=\ninerm   \scriptfont0=\sixrm
\scriptscriptfont0=\fiverm  \def\rm{\fam0\ninerm}%
\textfont1=\ninei   \scriptfont1=\sixi
\scriptscriptfont1=\fivei  \def\oldstyle{\fam1\ninei}%
\textfont2=\ninesy   \scriptfont2=\sixsy
\scriptscriptfont2=\fivesy
\textfont\itfam=\nineit  \def\it{\fam\itfam\nineit}%
\textfont\slfam=\ninesl  \def\sl{\fam\slfam\ninesl}%
\textfont\ttfam=\ninett  \def\tt{\fam\ttfam\ninett}%
\textfont\frakfam=\ninefrak \def\frak{\fam\frakfam\ninefrak}%
\textfont\bbfam=\ninebb  \def\Bbb{\fam\bbfam\ninebb}%
\textfont\bffam=\ninebf   \scriptfont\bffam=\sixbf
\scriptscriptfont\bffam=\fivebf  \def\bf{\fam\bffam\ninebf}%
\abovedisplayskip=10pt plus 2pt minus 6pt
\belowdisplayskip=\abovedisplayskip
\abovedisplayshortskip=0pt plus 2pt
\belowdisplayshortskip=5pt plus2pt minus 3pt
\smallskipamount=2pt plus 1pt minus 1pt
\medskipamount=4pt plus 2pt minus 2pt
\bigskipamount=10pt plus4pt minus 4pt
\setbox\strutbox=\hbox{\vrule height 7pt depth 2pt width 0pt}%
\normalbaselineskip=10pt \normalbaselines
\rm}


\def\pagewidth#1{\hsize= #1}
\def\pageheight#1{\vsize= #1}
\def\hcorrection#1{\advance\hoffset by #1}
\def\vcorrection#1{\advance\voffset by #1}

\newif\iftitlepage   \titlepagetrue               
\newtoks\titlepagefoot     \titlepagefoot={\hfil} 
\newtoks\otherpagesfoot    \otherpagesfoot={\hfil\tenrm\folio\hfil}
\footline={\iftitlepage\the\titlepagefoot\global\titlepagefalse
           \else\the\otherpagesfoot\fi}

\font\extra=cmss10 scaled \magstep0
\setbox1 = \hbox{{{\extra R}}}
\setbox2 = \hbox{{{\extra I}}}
\setbox3 = \hbox{{{\extra C}}}
\setbox4 = \hbox{{{\extra Z}}}
\setbox5 = \hbox{{{\extra N}}}

\def\RRR{{{\extra R}}\hskip-\wd1\hskip2.0
   true pt{{\extra I}}\hskip-\wd2
\hskip-2.0 true pt\hskip\wd1}
\def\Real{\hbox{{\extra\RRR}}}    

\def\CCC{{{\extra C}}\hskip-\wd3\hskip 2.5 true pt{{\extra I}}
\hskip-\wd2\hskip-2.5 true pt\hskip\wd3}
\def\Complex{\hbox{{\extra\CCC}}\!\!}   





\def\R{{\Real}}
\def\C{{\Complex}}

\def\frac#1#2{{#1\over#2}}

\def\({\left(}
\def\){\right)}
\def\<{\langle}
\def\>{\rangle}

\def\pmb#1{\setbox0=\hbox{$#1$}%
   \kern-.025em\copy0\kern-\wd0
   \kern.05em\copy0\kern-\wd0
   \kern-.025em\raise.0433em\box0 }


\def\abstract#1{{\parindent=30pt\narrower\noindent\ninepoint\openup
2pt #1\par}}


\newcount\notenumber\notenumber=1
\def\note#1
{\unskip\footnote{$^{\the\notenumber}$}
{\eightpoint\openup 1pt #1}
\global\advance\notenumber by 1}


\global\newcount\secno \global\secno=0
\global\newcount\meqno \global\meqno=1
\global\newcount\appno \global\appno=0
\newwrite\eqmac
\def\romappno{\ifcase\appno\or A\or B\or C\or D\or E\or F\or G\or H
\or I\or J\or K\or L\or M\or N\or O\or P\or Q\or R\or S\or T\or U\or
V\or W\or X\or Y\or Z\fi}
\def\eqn#1{
        \ifnum\secno>0
            \eqno(\the\secno.\the\meqno)\xdef#1{\the\secno.\the\meqno}
          \else\ifnum\appno>0
            \eqno({\rm\romappno}.\the\meqno)\xdef#1{{\rm\romappno}.\the=
\meqno}
          \else
            \eqno(\the\meqno)\xdef#1{\the\meqno}
          \fi
        \fi
\global\advance\meqno by1 }


\global\newcount\refno
\global\refno=1 \newwrite\reffile
\newwrite\refmac
\newlinechar=`\^^J
\def\ref#1#2{\the\refno\nref#1{#2}}
\def\nref#1#2{\xdef#1{\the\refno}
\ifnum\refno=1\immediate\openout\reffile=refs.tmp\fi
\immediate\write\reffile{
     \noexpand\item{[\noexpand#1]\ }#2\noexpand\nobreak.}
     \immediate\write\refmac{\def\noexpand#1{\the\refno}}
   \global\advance\refno by1}
\def\semi{;\hfil\noexpand\break ^^J}
\def\nl{\hfil\noexpand\break ^^J}
\def\refn#1#2{\nref#1{#2}}

\def\ann#1#2#3{{\it Ann. Phys.} {\bf {#1}} (19{#2}) #3}
\def\cmp#1#2#3{{\it Commun. Math. Phys.} {\bf {#1}} (19{#2}) #3}

\def\jmp#1#2#3{{\it J. Math. Phys.} {\bf {#1}} (19{#2}) #3}

\def\ijtp#1#2#3{{\it Int.  J.  Theor.  Phys.} {\bf {#1}} (19{#2}) #3}

\def\plA#1#2#3{{\it Phys.  Lett.} {\bf {#1}A} (19{#2}) #3}

\def\prD#1#2#3{{\it Phys.  Rev.} {\bf D{#1}} (19{#2}) #3}

\def\prp#1#2#3{{\it Phys.  Rep.} {\bf {#1}C} (19{#2}) #3}


{

\refn\Berry
{M.V. Berry,
{\sl Adv. Phys.} {\bf 25} (1976) 1}

\refn\MaF
{V.P. Maslov and M.V. Fedoriuk, \lq\lq Semiclassical
Approximation in Quantum \nl
Mechanics\rq\rq, D. Reidel Publ.,
London, 1981}

\refn\Schulman
{L.S. Schulman, \lq\lq Techniques and Applications of
Path Integration\rq\rq, John Wiley \& Sons, New York, 1981}

\refn\S
{L.S. Schulman, in \lq\lq Functional Integration
and its Applications\rq\rq, A.M. Arthurs, ed.,
Clarendon Press, Oxford, 1975}

\refn\DM
{C. DeWitt-Morette, \ann{97}{76}{367}}

\refn\DMMN
{C. DeWitt-Morette, A. Maheshwari and B. Nelson,
\prp{50}{79}{256}}

\refn\DMNZ
{C. DeWitt-Morette, B. Nelson and T.-R. Zhang,
\prD{28}{83}{2526}}

\refn\DV
{G. Dangelmayr and W. Veit,
\ann{118}{79}{108}}

\refn\FH
{R.P. Feynman and A.R. Hibbs,
\lq\lq Quantum Mechanics and Path Integrals\rq\rq,
McGraw-Hill, New York, 1965}

\refn\GY
{I.M. Gel'fand and A.M. Yaglom, \jmp{1}{60}{48}}

\refn\LS
{S. Levit and U. Smilansky,
{\sl Proc. Amer. Math. Soc.} {\bf 21} (1977) 299;
\ann{103}{77}{198}}

\refn\HMTT
{K. Horie, H. Miyazaki, I. Tsutsui and S. Tanimura,
{\sl Quantum Caustics for Systems with Quadratic Lagrangians},
KEK Preprint 98-161, hep-th/9810156}

\refn\CH
{R. Courant and D. Hilbert,
\lq\lq Methods of Mathematical Physics\rq\rq,
Interscience Publishers, New York, 1953}

\refn\MF
{P.M. Morse and H. Feshbach,
\lq\lq Methods of Theoretical Physics\rq\rq,
McGraw-Hill Book Company, New York, 1953}

\refn\Milnor
{J. Milnor, \lq\lq Morse Theory\rq\rq,
Princeton University Press, Princeton, 1963}

\refn\Souriau
{J.-M. Souriau, in \lq\lq Group Theoretical Methods
in Physics\rq\rq, A. Janner, T. Janssen and M. Boon, eds.,
Lecture Notes in Physics, {\bf 50}, Springer-Verlag, Berlin, 1976}

\refn\Horv
{P.A. Horv\'{a}thy, \ijtp{13}{79}{245}}

\refn\C
{B.K. Cheng, \plA{101}{84}{464}}

\refn\H
{G.A. Hagedorn, \cmp{71}{80}{77}}

\refn\HMT
{K. Horie, H. Miyazaki and I. Tsutsui,
{\sl Quantum Caustics for Systems with Quadratic Lagrangians
in Multi-Dimensions}, in preparation}







}

\def\ve{\vfill\eject}




\pageheight{23cm}
\pagewidth{14.8cm}
\hcorrection{0mm}
\magnification= \magstep1
\def\bsk{%
\baselineskip= 16.8pt plus 1pt minus 1pt}
\parskip=5pt plus 1pt minus 1pt
\tolerance 6000


\null

{
\leftskip=100mm
\hfill\break
KEK Preprint 98-195
\hfill\break
\par}

\smallskip
\vfill
{\baselineskip=18pt

\centerline{\bigbold
Quantum Caustics in the Gaussian Slit Experiment}

\vskip 30pt

\centerline{
\smc
Kenichi Horie,
\quad
Hitoshi Miyazaki
\quad {\rm and} \quad
Izumi Tsutsui\note
{E-mail:\quad tsutsui@tanashi.kek.jp}
}

\vskip 5pt

{
\baselineskip=13pt
\centerline{\it
Institute of Particle and Nuclear Studies}
\centerline{\it
High Energy Accelerator Research Organization (KEK),
Tanashi Branch}
\centerline{\it Tokyo 188-8501, Japan}
}

\vskip 15pt

\centerline{
\smc Shogo Tanimura\note
{E-mail:\quad tanimura@kuamp.kyoto-u.ac.jp}
}

\vskip 5pt

{
\baselineskip=13pt
\centerline{\it
Department of Applied Mathematics and Physics}
\centerline{\it
Graduate School of Informatics, Kyoto University}
\centerline{\it Kyoto 606-8501, Japan}
}

\vskip 60pt

\abstract{%
{\bf Abstract.}\quad
We study classical and quantum caustics for systems with
quadratic Lagrangians of the form
$L = {1 \over 2} \dot x^2 - {1 \over 2} 
\lambda(t) x^2 - \mu(t) x$.
After deriving the transition amplitude 
on caustics in a closed form, 
we consider the Gaussian slit experiment
and point out that the focusing
around caustics is stabilized 
against initial momentum fluctuations by quantum effect.
}

\bigskip
{\ninepoint
PACS codes: 02.30.Hg; 03.65.-w; 03.65.Sq \hfill\break
}


\pageheight{23cm}
\pagewidth{15.7cm}
\hcorrection{-1mm}
\magnification= \magstep1
\def\bsk{%
\baselineskip= 15pt plus 1pt minus 1pt}
\parskip=5pt plus 1pt minus 1pt
\tolerance 8000
\bsk

\ve

\secno=1 \meqno=1


\noindent{\bf 1. Introduction.}
When a family of classical paths focuses, the envelope
of the trajectories forms a focal region called caustics.
The non-existence of paths or a coalescence of trajectories
are interesting classical phenomena pertaining to caustics,
which have been studied extensively in various fields of physics
(see, {\it e.g.,} [\Berry]).
Their consequences on quantum physics are, however, somewhat
involved.  This is mainly due to the fact that the 
powerful and commonly used 
semiclassical approximation scheme [\MaF, \Schulman] 
breaks down when caustics occur, since in this approach 
the transition amplitude is obtained by summing up
fluctuations around (a finite number of) 
classical paths which are assumed to exist.
So far there have appeared numerous works on the problem
of quantum caustics.  On one hand, a generalized
prescription for the semiclassical approximation in the
path-integral framework has been devised [\S, \Schulman],
and on the other, a detailed quantum analysis for caustics
in the presence of higher order terms (in the semiclassical
expansion) --- which smear out and so 
defuse the singularity at caustics --- 
has been pursued [\S, \DM] 
(see also [\Schulman, \DMMN, \DMNZ, \DV]).

In this letter, we wish to provide a simple method to
derive a closed form of the transition kernel 
on caustics in the semiclassical approximation.  We also
mention some novel quantum aspects observed on and around
caustics, namely, the concentration of amplitude on 
the focal point, observed
in the Gaussian slit (gedanken-)experiment [\FH].
For simplicity we restrict ourselves to one dimensional
systems without higher order terms
focusing on systems 
governed by the quadratic Lagrangian of the form%
\note{Note that any Lagrangian at most quadratic in position
and velocity can be brought into this form by partial
integration.  We use the dot to denote time derivative
$\dot x = dx/dt$, and put the mass $m$ of the particle unity.}
$$
  L[x]  = {1 \over 2}\dot{x}^2   
     -{1 \over 2}\lambda(t)\,x^2
     -\mu(t)\,x\ ,
\eqn\lagrange
$$
for which the semiclassical approximation is known to be exact.
The transition amplitude between the two given endpoints
$a$, $b \in \R$ during the interval $[0, T]$ reads [\GY, \LS]
$$
K(b,T;a,0)
       = \left( {{i}\over{2\pi\hbar}}
               \left\vert{ {\partial^2 I[x_{\rm cl}]}
                          \over{\partial a \partial b}
               }\right\vert
         \right)^{1\over 2}\,
        e^{ {i\over\hbar}I[x_{\rm cl}]
           -{{i\pi}\over 2} m(\lambda) },
\eqn\semiexact
$$
where $m(\lambda)$ is the Morse index of the harmonic
potential $\lambda$ and
$I[x_{\rm cl}] = \int_0^T dt\, L[x_{\rm cl}]$ is
the action evaluated for the classical path $x_{\rm cl}(t)$
connecting the two endpoints $a$, $b$.
Clearly, when the classical path
ceases to exist, the prefactor 
$\vert\partial^2 I[x_{\rm cl}]/
\partial a \partial b \vert$ becomes singular and so does 
the transition kernel (\semiexact).
We shall see that the kernel at the singularity, {\it i.e.},  
on caustics, can nevertheless be obtained
in the modified prescription [\Schulman] by making use
of the unitarity relation of the kernel, which improves
the previous treatment [\DM] based on an abstract setting.    
Further, 
our analysis of the Gaussian slit experiment reveals that
the quantum effect
suppresses the susceptibility of the focal concentration
against initial momentum fluctuations.

\medskip

\secno=2 \meqno=1
\noindent{\bf 2. Classical caustics.}
Let us first recapitulate the classical caustics 
(for a fuller account, see [\HMTT]).
The equation of motion derived from (\lagrange) 
and satisfied by the classical
solution $x_{\rm cl}$ takes the form,
$$
 A_{\lambda}\, x_{\rm cl}(t) = \mu(t)\ , \qquad  \hbox{where}
\quad
  A_{\lambda} :=%
  - \left[ {{d^2}\over{dt^2}} + \lambda(t) \right] \ .
\eqn\motionfull
$$
If we let $x_{\rm cl}(p,t)$
be the solutions  
characterized by the initial momenta $\dot x(p, 0) = p$, then
the Jacobi field 
$
J(p,t) := \partial x_{\rm cl}(p,t)/\partial p
$ 
is a solution of the homogeneous part of 
the equation of motion (\motionfull),
$$
 A_{\lambda}\, J(p, t) = 0\ ,
\eqn\motion
$$
satisfying $J(p, 0)=0$.  Since 
the Jacobi field so defined 
gives a measure for the variation at later times 
under initial momentum fluctuations, we see that caustics
(at $t = T$) occur when its final value vanishes $J(p, T) = 0$.  
Note that, on account of the relation
$\partial^2 I[x_{\rm cl}]/\partial a \partial b = - 1/J(p, T)$,
the classical caustics implies quantum caustics 
as well, as seen in (\semiexact).
Let $u$, $v$ be two 
linearly independent solutions of the
homogeneous equation (\motion)
subject to the initial conditions
$ u(0) = 0 $, $ \dot u(0) = 1 $ and
$ v(0) = 1 $, $ \dot v(0) = 0 $.
(In terms of the Jacobi field, we have
$u(t) = J(p, t)/\dot J(p, 0)$.)
Then the general solution of the
full equation (\motionfull) is given by
$$
x(t) = \alpha\, v(t) + \beta\, u(t) + s(t) \ ,
\eqn\gensolution
$$
where $s$ is a special solution of (\motionfull)
satisfying $s(0) = 0$. 
The constants $\alpha$, $\beta$ are determined from 
the initial position $x(0)$ and velocity
$\dot x(0)$, respectively. 

When considering the quantum 
transition amplitude in semiclassical expansion, it is
important to know whether to a given Dirichlet boundary
condition
$$
  x(0) = a, \qquad x(T) = b
\eqn\vab
$$
there exists a classical trajectory.
Suppose first there is no caustic at $t=T$, that is,
the Jacobi field $J(p,t)$ (and thus also $u(t)$) does
not vanish at $t=T$.  Then, by choosing $\alpha = a$ and
$\beta = (b - a\, v(T) - s(T))/u(T)$ in the
general solution (\gensolution), one obtains the desired
trajectory which is clearly unique.
However, when caustics occur at $t=T$ ({\it i.e.,}
$J(p, T) = 0$), then the final position
$x(T)$ turns out to be independent of
the initial velocity $\dot x(0)$, and while
$\alpha = a$, $\beta$ is left arbitrary since $u(T) = 0$.
Thus whatever the initial velocity 
may be, all paths starting from $x(0) = a$ ends
at the final point
$$
  x(T) = v(T)\cdot a + s(T)\ .
\eqn\klambda
$$
If the given value $b$ does not match this unique value,
then there is no classical trajectroy satisfying the
Dirichlet boundary condition (\vab).

This phenomenon can be considered as a special case of
caustics in geometric optics (see {\it e.g.}, [\Schulman]),
and when it occurs we shall call the potential $\lambda$
{\it critical}, otherwise {\it non-critical}.
For those critical potentials, it is convenient to introduce 
the constant
$$
  k(\lambda) := {{v(T)}\over{v(0)}}\ ,
\eqn\kdef
$$
where here $v$ can be any solution of the homogeneous part
(\motion) with $v(0) \ne 0$.  It is easy to show
that $k(\lambda)$ is indeed independent of $v$, and in view
of (\klambda), it gives the stretching factor during the period
$[0,T]$. The final point $x(T)$ in (\klambda) is called the
{\it conjugate point} to $a$ or {\it focal point}.

If $\lambda$ is critical, then besides the
stretching factor $k(\lambda)$, a further characteristics
is given by the Morse index, which we now briefly outline.
Let $I[x]$ be the action functional for paths $x$ with
fixed boundary condition (\vab).
The second variation of the action $\delta^2 I$ around
a given classical path $x_{\rm cl}(t)$ (which in case of
critical potential $\lambda$ is assumed to exist) is a
symmetric bilinear functional of the perturbations around
$x_{\rm cl}$, and its kernel is given by the operator
$A_\lambda$ in (\motionfull).  From the consideration of 
Sturm-Liouville problem the operator $A_\lambda$ 
is known to possess a complete set
of orthonormal eigenfunctions $u_n(t)$ [\CH, \MF]:
$$
  A_{\lambda} u_{n}(t) =
  -\left[ {{d^2}\over{dt^2}} + \lambda(t) \right]
  u_{n}(t) = E_n\, u_{n}(t)\ ,
\eqn\eigen
$$
with
$$
  u_n(0) = u_n(T) = 0\, ; \qquad
  \int_0^T dt\, u_n(t)\, u_m(t) = \delta_{nm} .
\eqn\usn
$$
The index of the bilinear functional (Hessian)
$\delta^2 I$ is given by the number of negative eigenvalues
of eigenvalue problem (\eigen) for non-degenerate $A_\lambda$
for which no zero mode exists.
It characterizes the type of the saddle point of the action
functional $I[x]$ at the extremum path $x_{\rm cl}$.
(In this letter, we assume the eigenvalues to be
bounded from below, which is for example the case
when $\lambda$ is bounded from above.)
The index is known to be equal to the {\it Morse index}
which is the number of zeros of the Jacobi field on the
half-open interval $(0,T]$ [\Milnor] and denoted by
$m(\lambda)$.
Note that the bilinear functional $A_\lambda$ 
becomes degenerate if there arises a zero mode 
solution of (\eigen), a $u_m$ with $E_m = 0$ for some $m$, 
{\it i.e.}, when
caustics occur.
In this case the Morse index gives the number of negative
eigenvalues plus one, and accordingly the index of $A_\lambda$
may be extended even to this degenerate case by saying that
it is given by the number of modes with $E_n \le 0$.

\medskip

\secno=3 \meqno=1
\noindent{\bf 3. Quantum caustics.}
We now study the quantum dynamics of the system (\lagrange).
Let $I[x]$ be the action evaluated along a path fulfilling
the boundary condition (\vab).  Corresponding to 
this boundary condition (\vab),
the transition amplitude in the path-integral is given by
$$
        K (b,T;a,0)
    = \int_{x(0)=a}^{x(T)=b} {\cal D} x \, e^{{i\over\hbar} I[x]}.
\eqn\path
$$
In carrying out the path-integration, we need to
take into account the fact that, when caustics occur,
there may not exist a classical solution
that respects the given boundary condition.
The general case, including the caustic one,
may be handled by the following procedure [\Schulman, \S].

Let $\bar x_{\rm cl}$ be a classical solution of
the equation of motion (\motionfull) with the initial 
value $\bar x_{\rm cl}(0) = a$, and let $c:=\bar x_{\rm cl}(T)$
be the endpoint of the solution. 
Note that in general $c$ may not be equal
to $b$, but if there is a solution having $b$ as its
endpoint, then we shall omit the bar and denote this
special solution by $x_{\rm cl}$. 
Such a solution exists
if $\lambda$ is non-critical, whereas if $\lambda$ is
critical the endpoint is determined uniquely
$c = k(\lambda)\,a + s(T)$, see (\klambda) and (\kdef).
Now, let us introduce a fixed, smooth function $\rho(t)$
satisfying
$ \rho(0) = 0$ and $\rho(T) = b - c $.
Then any path obeying the boundary condition (\vab)
may be decomposed as
$$
        x(t) = \bar x_{\rm cl}(t) + \rho(t) + \eta(t)\ ,
\eqn\xdecomp
$$
with $\eta(t)$ being the function representing
the fluctuations vanishing at both of the ends
$ \eta(0) = \eta(T) = 0$.
With the help of the orthonormal eigenfunctions in (\eigen)
and (\usn) it may be expanded as $ \eta(t) = \sum_n a_n u_n(t) $.
Using this and the decomposition (\xdecomp) we find by partial
integrations
$$
\eqalign{
        I[ \bar x_{\rm cl} + \rho + \eta ]
        &=
        I[ \bar x_{\rm cl} ] + I[ \rho ]\vert_{\mu = 0}
        +  \dot{\bar x}_{\rm cl}(T) \rho(T) 
        \cr
        &\qquad +
        \frac{1}{2} \sum_n E_n a_n^2
        + \rho(T) \sum_n a_n \dot{u}_n (T)
        + \sum_n E_n a_n \int_0^T dt \, \rho(t) u_n(t).
}
\eqn\calculus
$$
In order to perform the path-integral (\path)
we change the integral variables
${\cal D}x = {\cal D}\eta =  {\cal N}\prod_n da_n$
where $ {\cal N} $ is a Jacobian factor.
Further, we replace $E_n$ with 
$E_n + i\epsilon$, $\epsilon > 0$
infinitesimal, such that the integrations over $a_n$
become all Gaussian even if an eigenvalue $E_m$ should vanish.
With these preparations one obtains
$$
        K(b,T;a,0) 
        = {\cal N}
        \left[ \prod_n (E_n + i \epsilon) \right]^{ -\frac12 }\,
        e^{ {i\over\hbar} \Phi(b, a; \lambda) }\ ,
\eqn\genkernel
$$
where
$$
\eqalign{
\Phi(b, a; \lambda)
&:= I[ \bar{x}_{\rm cl} ] + I[ \rho ]\vert_{\mu = 0} +
\dot{\bar{x}}_{\rm cl}(T) \rho(T) \cr
&\qquad - \frac{1}{2} \sum_n \frac{1}{ (E_n  + i \epsilon)}
                \left\{
                        \rho(T)\, \dot{u}_n (T)
        + (E_n + i \epsilon) \int_0^T dt \, \rho(t)\, u_n(t)
                \right\}^2\ .
}
\eqn\genphase
$$
The kernel formula (\genkernel) is valid for any potential
$\lambda$.
Now for $\lambda$ non-critical, each $E_n \ne 0$, and
in the limit $\epsilon \rightarrow 0$ the kernel
(\genkernel) reduces to
$$
        K(b,T;a,0) =
        {\cal N}
        \left[ \prod_n E_n \right]^{ -\frac12 }
        e^{{i\over\hbar} I[ x_{\rm cl}] }\ ,
\eqn\kernelnonc
$$
where $\rho(t)$ has been set identically zero in (\genphase)
owing to the existence of the solution $x_{\rm cl}$.
The obtained kernel (\kernelnonc) can be shown
[\GY, \LS] to be equivalent to (\semiexact).

If, on the other hand, $ \lambda(t) $ is critical,
then $ E_m = 0 $ for some $ m $, and the integration over
$a_m$ yields in the limit $\epsilon \rightarrow 0$
the delta-function $\delta( \rho(T)\, \dot{u}_m (T) )$
on account of the identity,
$
\lim_{\epsilon \rightarrow 0}
\frac{1}{ \sqrt{2 \pi \epsilon} } \,
        e^{ - {x^2}/{2\epsilon}}  = \delta(x)\ .
$
Hence, for critical $ \lambda(t) $ the kernel formula
(\genkernel) reduces to 
$$
        K(b,T;a,0) =
        \sqrt{\frac{2 \pi}{i}} {\cal N}
        \left[ \prod_{n \ne m} E_n\right]^{ -\frac12 }
        \delta( \rho(T) \,\dot{u}_m (T) )\, 
        e^{{i\over\hbar}I[x_{\rm cl}]}\ .
\eqn\ckera
$$
In (\ckera) we have set $\rho(t) = 0$ in the phase part of
the kernel (and hence we put 
$I[\bar x_{\rm cl}] = I[x_{\rm cl}]$) , 
which is allowed due to 
the delta-function and 
the fact $\dot{u}_m (T) \ne 0$. 
(This follows since both
${u}_m (T) = 0$ and $\dot{u}_m (T) = 0$ would imply $u_m = 0$ 
identically.)  We note that
$$
\rho(T) = b - k(\lambda)a - s(T)
\eqn\valrho
$$
and, therefore, the kernel (\ckera) clearly shows that
classically forbidden transitions to non-conjugate points,
$b \ne k(\lambda)a + s(T) $ from $a$, remains forbidden
even quantum mechanically.

The remarkable fact in the case of caustics
is that one can express the transition kernel (\ckera)
in terms of the stretching factor $k(\lambda)$
and the Morse index $m(\lambda)$.
To see this, let us write the kernel (\ckera) in the polar 
form,
$$
K(b,T;a,0) =  R(T)\, \delta ( \rho(T) )\, e^{ i \Theta (T) }.
\eqn\modulpath
$$
Noting that the transition kernel $K(b,T;a,0)$ can be
expressed as $\langle b |\, \widehat U(T,0)\, | a \rangle$
in terms of time evolution operator $\widehat U(T,0)$, 
the unitarity relation,
$$
        \int db\, K^* (b,T;c,0) \, K (b,T;a,0) =
        \int db\, \langle c | \,
        \widehat U^\dagger(T,0)\,  | b \rangle
        \langle b | \, \widehat U(T,0)\,  | a \rangle
        = \delta (a-c),
\eqn\unitarity
$$
determines the modulus $R(T)$ to be
$ R(T) = \sqrt{\vert k(\lambda)\vert} $.
On the other hand, since the Morse index gives
the number of negative modes plus one, we have
$
        [\prod_{n \ne m} E_n
         ]^{- \frac12}
        =
        [\prod_{n \ne m} | E_n |
        ]^{- \frac12}
        e^{- {{i\pi}\over 2} (m( \lambda ) - 1) }
$.
Thus the phase part is given by
$\Theta (T) = 
{1\over\hbar}I[x_{\rm cl}] - {\pi\over 2}(m(\lambda) - 1)$.
By combining these we find that the transition kernel
(\ckera) takes the closed form 
in terms of the stretching factor and Morse index,
$$
        K(b,T;a,0) =
        \sqrt{|k(\lambda)|}\,
        \delta\bigl( b - k(\lambda) a - s(T) \bigr)\,
        e^{ {i\over\hbar} 
            I[ x_{\rm cl} ] - {{i\pi}\over 2} m(\lambda) },
\eqn\ckerb
$$
up to an overall constant of unit modulus.
This final form agrees with the expression obtained
earlier based on a different construction of the path-integral
[\DM].

As application, take, for example,
the forced harmonic oscillator
given by $\lambda(t) = \omega^2$ and
$\mu(t) = - f(t)$.  A special solution satisfying
$s(0) = 0$ is then
$$
s(t) = {1 \over \omega} \int^t_0 dt'\,
\sin \omega(t - t')\, f(t').
\eqn\spesol
$$
For the present case,
caustics occur at $\omega = {{n\pi}\over T}$ with
$n = 1, 2, \ldots$, where we have
$k(\lambda) = (-1)^n$ and $m(\lambda) = n$, respectively.
The classical trajectory starting from $x_{\rm cl}(0) = a$
and ending up at the point (\klambda) possesses the action,
$$
I[ x_{\rm cl} ] = a \int^T_0 dt\, \cos \omega t\, f(t)
- {1\over\omega} \int^T_0 dt \int^t_0 dt'\,
\cos \omega t\,\sin \omega t'\, f(t)\,f(t')\ .
\eqn\clfho
$$
Hence, in particular, for constant $f$
the kernel (\ckerb) reduces to
$$
K(b,T;a,0) = \delta
\Bigl(
b - (-1)^n a - \{ 1 - (-1)^n \} f/\omega^2
\Bigr)\,
        e^{ {i\over\hbar} f^2T/2\omega^2 - in\pi/ 2 }.
\eqn\kernfho
$$
This formula coincides with the previous results 
[\Souriau, \Horv, \C] obtained
by other indirect means\note{The result in [\C]
obtained for $f(t) \ne 0$ is marred by an error in the phase.}
for the (forced) harmonic oscillator.

\medskip

\secno=4 \meqno=1
\noindent{\bf 4. Gaussian slit experiment.}
For $\lambda$ non-critical but close to a critical
$\bar\lambda$, one expects that there will be a
concentration in the transition amplitude around the
focal point $b = k(\bar\lambda) a + s(T)$ conjugate
to a given initial point $a$.
We analyze how the concentration takes place quantum
mechanically, based on the Gaussian slit (gedanken)
experiment [\FH].

Let there be an apparatus which emits a particle from
the origin $x = 0$ at time $t = - \tau$.  To get a Gaussian
distribution at $t = 0$, we place a \lq Gaussian
slit' centered at $x = a$ with effective width (variance)
$\sigma_0$, which is given by the Gaussian transmission factor
$$
G(x - a; \sigma_0) = N\,
\exp\left\{ -{{(x - a)^2}\over{4\sigma_0^2}} \right\}.
\eqn\gaussfactor
$$
The wave function $\psi(x, 0)$ of the particle at $t = 0$
is then furnished by the product of the free particle
kernel
$
K_0(x, 0; 0, -\tau)
= \sqrt{1/(2\pi i \hbar \tau)}\, e^{i x^2/2\hbar\tau }
$
and the transmission factor.  From the normalization condition
$\int dx \, \vert \psi(x, 0) \vert^2 = 1$ we determine the
constant $N$ and obtain
$$
\psi(x, 0) =  G(x-a; \sigma_0)\,  K_0(x, 0; 0, -\tau)
= {1\over{(2\pi\sigma_0^2)^{1\over 4}}}\,
\exp\left\{
-{{(x - a)^2}\over{4\sigma_0^2}}
+ {{i x^2 }\over{2\hbar \tau}}
\right\}\ ,
\eqn\initialwave
$$
which has the average momentum
$p = \int dx\, \psi^*(x, 0)
\left(-i\hbar {d\over{dx}}\right) \psi(x, 0)
= a/\tau$.  Using the kernel (\semiexact), the 
wave function at $t = T$ is then given by 
$$
\eqalign
\psi(y, T) = \int dx\, K(y, T; x, 0)\, \psi(x, 0).
\eqn\finalwave
$$

To simplify the ongoing discussion let us assume
first $\mu = 0$ in (\lagrange). Then, the classical
action $I[x_{\rm cl}](y, T; x, 0)$ is a quadratic
polynomial homogeneous in $x$ and $y$, because we have
$
I[x_{\rm cl}](cy, T; cx, 0) = I[c x_{\rm cl}]
= c^2 I[x_{\rm cl}] = c^2 I[x_{\rm cl}](y, T; x, 0)
$ for any constant $c$. Thus, with 
$A$, $B$ and $C$ some functions of $T$, we may write
$$
I[x_{\rm cl}](y, T; x, 0) = A x^2 + B xy + C y^2\ .
\eqn\claction
$$
Then it is straightforward to see that the wave function
(\finalwave) at time $t = T$ has the form
$$
\psi(y, T) =
{1\over{(2\pi\sigma^2(T))^{1\over 4}}}
\exp\left\{
-{ {\left( y - x_{\rm cl}(T) \right)^2}
  \over{4\sigma^2(T)}}
+ i\, ({\rm phase})
\right\},
\eqn\finalwf
$$
with
$$
x_{\rm cl}(T) = - {1 \over B} (2 a A + p),
\eqn\gencl
$$
and
$$
\sigma(T) = \sigma_0
\left\{
\left({{x_{\rm cl}(T)}\over{a}}\right)^2
+
\left({{\hbar}\over{2\sigma_0^2 B}}
\right)^2
\right\}^{1\over 2}.
\eqn\genvar
$$
The second term in (\genvar) represents the quantum effect
whereas the first term is just the classical variance,
since the variation in the initial position
$\Delta x_{\rm cl}(0) = \sigma_0$ around the center
$x_{\rm cl}(0) = a$ for the classical path results in
the final variation
$\Delta x_{\rm cl}(T) = \sigma_0 \vert x_{\rm cl}(T)\vert/a
= \sigma_0 \vert k(\lambda)\vert$.
(Note that the initial momentum $p$ is linearly dependent
on the initial position where the path goes through.)
Thus, as one expects, the quantum effect always enhances
the spread of the Gaussian distribution.

Viewed as a function of the initial variance $\sigma_0$,
the final variance $\sigma(T)$ attains its minimum,
{\it i.e.}, the highest concentration of intensity
$$
\sigma_{\rm min}(T) =
\left\vert
{{\hbar x_{\rm cl}(T)}\over{a B}}
\right\vert^{1\over 2}
\qquad
\hbox{at}
\quad
\sigma_0 = \left\vert
{{\hbar a }\over{2 B x_{\rm cl}(T)}}
\right\vert^{1\over 2},
\eqn\geninimini
$$
which is precisely the point where the quantum effect
matches the classical variance.
It then follows from (\gencl) and (\geninimini) that, 
ideally, the infinite concentration 
$\sigma_{\rm min}(T) = 0$
takes place at $B = \pm \infty$ or at $p = -2aA$.
The former case corresponds to the expected caustics,
which occur when 
the variance (\genvar) becomes purely classical
$\sigma(T) = \sigma_0 \vert k(\lambda) \vert$ for which
the infinite concentration is obtained by letting 
$\sigma_0 \rightarrow 0$.
By contrast, in the latter case the variance becomes
purely quantum mechanical, as all paths passing through
the slit coalesce toward the origin $x = 0$ at $t = T$.
In fact, this case is again caustics occurring during
the combined period $[-\tau, T]$ under the potential
$\lambda(t)$ which vanishes for $t < 0$, where
the infinite concentration at the
quantum level is achieved if we let
$\sigma_0 \rightarrow \infty$. 
It is obvious, however, that 
both of these concentrations in intensity are unstable,
since a small fluctuation in the parameters $B$
or $p$ will generally bring the variance to a large value.
Nevertheless, one could achieve a sharp peak in intensity
by adjusting the initial variance along with the parameters
according to (\geninimini).

An important quantity for characterizing the concentration
is the susceptibility of the variance against initial momentum
fluctuations.
Exposing the momentum dependence of the variance explicit
$\sigma(T) = \sigma(p, T)$, we use the following normalized
quantity for the susceptibility,
$$
S(p, T) :=
{a \over {\sigma_0}} {{\partial}\over{\partial p}}
\sigma(p, T).
\eqn\qjacobi
$$
The susceptibility of (\genvar) reads
$$
S(p, T) = \left\vert J(p, T) \right\vert
\left\{
1 +
\left(
{{\hbar a}\over{2 \sigma_0^2 B x_{\rm cl}(T)}}
\right)^2
\right\}^{-{1\over 2}},
\eqn\qjgen
$$
where $J(p, T) = - \left(\partial^2 I[x_{\rm cl}]
/\partial y \partial x \right)^{-1} = - 1/B$
is the Jacobi field for the classical action (\claction).
In the classical limit $\hbar \rightarrow 0$, the
susceptibility reduces to the (absolute value of the) Jacobi
field. In the two cases mentioned above in which an infinite
concentration can in principle be possible, the susceptibility
vanishes $S(p,T)=0$ and hence the Gaussian wave packet becomes
free from momentum fluctuations. What is interesting in
(\qjgen) is that the quantum effect suppresses the
susceptibility of the variation against initial momentum
fluctuations.

The above features of the Gaussian wave packet persist
even for $\mu \ne 0$, because then the classical action
$I[x_{\rm cl}](y, T; x, 0)$ acquires only linear and constant
terms in addition to the quadratic terms in (\claction) as
can be explicitly seen from the form of the general classical
solution derived after eq.\ (\vab). Accordingly, the time
evolution by the integral (\finalwave) is essentially
unchanged. Semiclassically, this may also be the case for
more general systems, not only for those with
quadratic Lagrangians we considered, in view of
the earlier study [\H] which suggests that these features
are a norm for a generic system in the limit $\hbar \rightarrow 0$.

As an example of the above discussion, consider again 
the harmonic oscillator $\lambda = \omega^2$ (but with $\mu = 0$).
>From the familiar classical action [\FH] one reads off that
$A = C = \omega/2\tan\omega T$ and $B = - \omega/\sin\omega T$.
With the center of the distribution
$x_{\rm cl}(T) = a\cos \omega T + (p/\omega) \sin \omega T$, 
one finds the variance, 
$$
\sigma(T) = \sigma_0
\left\{
\left( 
\cos \omega T + {p\over{a\omega}} \sin \omega T
\right)^2
+
\left({{\hbar\sin\omega T}\over{2\sigma_0^2\omega}}
\right)^2
\right\}^{1\over 2},
\eqn\variance
$$
possessing its minimum
$
\sigma_{\rm min}(T) =
\vert
\hbar x_{\rm cl}(T) \sin\omega T/a \omega
\vert^{1\over 2}
$ 
at
$
\sigma_0 = \vert
\hbar a \sin\omega T/2 \omega x_{\rm cl}(T)
\vert^{1\over 2}.
$
One thus sees that an infinite
concentration of the amplitude takes place either
at $\omega = {{n\pi}\over T}$ or 
$p = - a \omega/\tan \omega T$.

\medskip

\secno=5 \meqno=1
\noindent{\bf 5. Conclusion.}
When caustics occur in one dimension, 
only those paths connecting the
conjugate points are allowed classically, and the
intrinsic features of caustics are characterized by
the stretching factor $k(\lambda)$ and the Morse index
$m(\lambda)$.
In quantum mechanics the transition is still allowed
only between conjugate points, and for systems with
quadratic Lagrangians we derived the
path-integral kernel for the amplitude in a closed form,
expressed solely in terms of the stretching factor, the Morse
index, and the action of (any of the) solution paths.
As for the situation near caustics, the Gaussian slit
experiment showed that, although the variance itself
is enhanced at the quantum level, the susceptibility of the
variance of the packet against initial momentum fluctuations
is suppressed.
This was derived by considering a quantum analogue of the
Jacobi field, which reduces to the ordinary Jacobi field in
the classical limit.
Possible extensions of this work on quantum caustics,
{\it e.g.,} effects of higher order terms in the Lagrangian
on the quantum susceptibility or related problems in higher
dimensions, are currently under investigation [\HMT].

\ve
\baselineskip= 15.5pt plus 1pt minus 1pt
\parskip=5pt plus 1pt minus 1pt
\tolerance 8000
\vfill\eject

  \vfill\eject\immediate\closeout\reffile
  \centerline{{\bf References}}\bigskip\frenchspacing%
  \input refs.tmp\vfill\eject\nonfrenchspacing
\bye